\documentclass[cite,published]{epl}
\issue{76}{5}{2006}{774}{1 December 2006}%
\DOI{10.1209/epl/i2006-10358-3}

\newcommand{\Ghom}{\Gamma_{\mathrm{hom}}}
\newcommand{\Ginh}{\Gamma_{\mathrm{inh}}}
\newcommand{\Heff}{\mathcal{H}_{\mathrm{eff}}}

\title{Inhomogeneous losses and complexness of wave functions in chaotic cavities}
\shorttitle{Inhomogeneous losses and wave function complexness}

\author{D. V. Savin\inst{1,2} \and O. Legrand\inst{3} \and F. Mortessagne\inst{3}}
\shortauthor{D. V. Savin \textit{et al.}}

\recff{2 August 2006}{10 October 2006}%
\onlinepub{1 November 2006}

\institute{
\inst{1} Department of Mathematical Sciences, Brunel University - Uxbridge, UB8 3PH, UK\\
\inst{2} Fachbereich Physik, Universit\"at Duisburg-Essen - 45117 Essen, Germany\\
\inst{3} Laboratoire de Physique de la Mati\`ere Condens\'ee, CNRS UMR 6622
\\ Universit\'e de Nice-Sophia Antipolis - 06108 Nice cedex 2, France }

\begin{document}

\vspace*{-2ex}%
\maketitle

\begin{abstract}
In a two-dimensional microwave chaotic cavity ohmic losses located at the
contour of the cavity result in different broadenings of different modes. We
provide an analytic description and establish the link between such an
inhomogeneous damping and the complex (non-real) character of  biorthogonal
wave functions. This substantiates the corresponding recent experimental
findings of Barth\'elemy \textit{et al.} [Europhys. Lett. \textbf{70} (2005)
162].
\end{abstract}

Open wave-chaotic systems in the presence of energy losses (absorption) are
nowadays under intense experimental and  theoretical investigations, see
\cite{Kuhl2005r,Fyodorov2005r} for recent reviews as well as
\cite{Stoeckmann} for a general discussion. Most of the works concern the
case of uniform absorption which is responsible for homogeneous broadening
$\Ghom$ of all the modes (resonance states). However, in some experimentally
relevant situations like, e.g., complex reverberant structures
\cite{Lobkis2000i,Rozhkov2003} or even microwave cavities at room temperature
\cite{Barthelemy2005b,Barthelemy2005a} one should take into account also
localized-in-space losses which lead to an inhomogeneous part $\Ginh$  of the
widths which varies from mode to mode. As a result, the neighboring modes
experience nontrivial correlations due to interference via one and the same
decaying / dissipative environment that result in the complex-valued wave
functions of corresponding resonance states. Such a complexness may reveal
itself in long-range correlations of wave function intensity and current
density \cite{Brouwer2003} that were recently studied experimentally
\cite{Kim2005}. Following Refs.~\cite{Pnini1996,Lobkis2000i}, it is
convenient to measure the above mentioned complexness through a single
statistical parameter, namely the ratio
$\langle(\mathrm{Im}\psi)^2\rangle/\langle(\mathrm{Re}\psi)^2\rangle=q^2$ of
variances of the real and imaginary parts of the mode wave function $\psi$.
The modes are real ($q=0$) in the case of vanishing inhomogeneous losses and
become complex-valued when $\Ginh\neq0$ (the value of $\Ghom$ has no effect
on the mode complexity, as it will become clear later on).  The strong
experimental evidences in the favor of the intimate relation between $q$ and
$\Ginh$ were recently provided by Barth\'elemy \textit{et al.}
\cite{Barthelemy2005b}, who analyzed hundreds of resonances in a Sinai-like
chaotic microwave cavity and found a proportionality between these two
quantities. Here, we present a thorough analytic description of this problem.

It is instructive to start with the discussion of the physical picture of
the problematic. A typical experimental setup consists of a flat
(two-dimensional) cavity fed with microwaves through attached antennas or
waveguides which support $M_a$ scattering channels (propagating modes) in
total. These very channels are used to perform all the measurements and
their number is finite. Dissipation through ohmic losses at cavity
boundaries gives another reason to treat our system as open, since
boundaries may be viewed locally as distributed ``parasitic" equivalent
channels with \textit{ad hoc} impedances
\cite{Lewenkopf1992i,Brouwer1997ii}. One should distinguish, however,
between almost uniform ohmic attenuation at the cavity plates and localized
absorption at the cavity contour \cite{Barthelemy2005b,Barthelemy2005a}. The
number of ``parasitic" channels responsible for the former (``bulk")
mechanism can be naturally estimated as $M_b\sim(L/\lambda)^2$, while one
has typically $M_c\sim L/\lambda$ channels at the contour (where $L$ is the
characteristic length of the cavity and $\lambda$ denotes the wave length).
Both $M_b,M_c\gg1$ but their ratio $M_c/M_b\sim\lambda/L$ is parametrically
small and that will be essential for our consideration.

It is natural, therefore, to use the following model description of the
problem. According to the Hamiltonian approach to scattering
\cite{Verbaarschot1985,Sokolov1989,Fyodorov1997},  see also
\cite{Fyodorov2005r}, one can represent the scattering matrix in terms of
the effective non-Hermitian Hamiltonian $\Heff$ of the open system as
follows:
\begin{equation}\label{Stot}
S_{\mathrm{tot}} = 1 - i V^{\dagger}\frac{1}{E-\Heff}V\,, \quad\qquad %
\Heff = H - \frac{i}{2}VV^{\dagger}\,.
\end{equation}
The Hamiltonian $H$ of the closed chaotic system gives rise to $N$ levels
(eigenfrequencies) $\epsilon_n$ characterized locally in the relevant range
of the energy $E$ by the mean level spacing $\Delta$. Those are coupled to
all the open channels via the $N\times(M_a+M_b+M_c)$ matrix $V$ of coupling
amplitudes and, as a result, are converted into complex resonances
$\mathcal{E}_n=E_n-\frac{i}{2}\Gamma_n$, which are given by the poles of the
$S$-matrix. Accordingly, we propose to decompose $V=\{A,B,C\}$ into coupling
to $M_a$ antennas, $M_b$ ``bulk" and $M_c$ ``contour" parasitic channels to
separate explicitly different contributions to the widths $\Gamma_n$. The
total $(M_a+M_b+M_c)$-dimensional scattering matrix (\ref{Stot}) is, of
course, unitary\footnote{Unitarity  of the total $S$-matrix as well as
causality, $\Gamma_n>0$, are automatically provided by the factorized
algebraic structure of the anti-Hermitian part of $\Heff$
\cite{Sokolov1989}.}.   However, one can access experimentally only the
($M_a\times M_a$) subblock $S = 1 - i A^{\dagger}(E-\Heff)^{-1}A$ which is
subunitary.

Without loss of generality one can consider the case of preserved
time-reversal invariance (which was indeed the case of experiment
\cite{Barthelemy2005b}) when the coupling amplitudes $V$ are real and $H$ is
symmetric. It is convenient first to represent $\Heff$ in the eigenbasis of
its Hermitian part as follows:
\begin{equation}
(\Heff)_{nm}=\epsilon_n\delta_{nm}-\frac{i}{2}\left(\sum_{a=1}^{M_a}A_n^a
A_m^a + \sum_{b=1}^{M_b} B_n^b B_m^b + \sum_{c=1}^{M_c} C_n^c C_m^c\right).
\end{equation}
In a chaotic cavity, $H$ is commonly described in the framework of Random
Matrix Theory \cite{Stoeckmann}. A (real orthogonal) rotation that
diagonalizes the random Hermitian matrix $H$ transforms the (fixed) matrix
$V$ to Gaussian-distributed coupling amplitudes with zero means and following
covariances (we assume statistical independence of channels from different
classes) \cite{Verbaarschot1985,Sokolov1989}:
\begin{equation}\label{cov}
\langle A_n^a A_{n'}^{a'}\rangle = 2\kappa_a \frac{\Delta}{\pi}
\delta^{aa'}\delta_{nn'}\,, \quad
\langle B_n^b B_{n'}^{b'}\rangle =
2\kappa_b \frac{\Delta}{\pi} \delta^{bb'} \delta_{nn'}\,, \quad
\langle
C_n^c C_{n'}^{c'}\rangle = 2\kappa_c \frac{\Delta}{\pi} \delta^{cc'}
\delta_{nn'}\,.
\end{equation}
Coupling constants $\kappa$ determine transmission coefficients
$T=4\kappa/(1+\kappa)^2$ of the corresponding channels, so that $T\ll1$
($T=1$) stands for weak (perfect) coupling. The strong inequality $M_b\gg
M_c \gg1$ allows us to perform now the limit of a very large number of weak
fictitious bulk channels, $M_b\to\infty$ and $T_b\to0$ with
$M_bT_b\equiv2\pi\Ghom/\Delta$ being kept fixed \cite{Brouwer1997ii}, which
singles out the homogeneous absorption contribution. Indeed, by virtue of
the central limit theorem one may replace in $\Heff$ the sum
$\sum_{b=1}^{M_b} B_n^b B_m^b$ with its average value
$\sum_{b=1}^{M_b}\langle B_n^b B_m^b\rangle\equiv\Ghom\delta_{nm}$ in the
limit considered that yields
\begin{equation}
(\Heff)_{nm}=\left(\epsilon_n-\frac{i}{2}\Ghom\right) \delta_{nm} -
\frac{i}{2} \left(AA^{\mathrm{T}}+CC^{\mathrm{T}}\right)_{nm}\,,
\end{equation}
meaning that all
the levels acquire one and the same attenuation rate. Since $\Heff$ comes
into the scattering problem only as the resolvent $(E-\Heff)^{-1}$, uniform
absorption turns out to be operationally  equivalent to a pure imaginary
shift of the scattering energy $E\to E+\frac{i}{2}\Ghom\equiv E_{\gamma}$
\cite{Brouwer1997ii,Savin2003i}. Thus the physical scattering matrix
acquires the following form:
\begin{equation}\label{S}
S = 1 - i A^{\mathrm{T}}\frac{1}{E_{\gamma}-\widetilde\Heff}A\,, \quad\qquad
\widetilde\Heff = H - \frac{i}{2}(AA^{\mathrm{T}}+CC^{\mathrm{T}})\,.
\end{equation}
A representation similar to (\ref{S}) was used in
\cite{Rozhkov2003,Rozhkov2004} to study statistics of transmitted power in
dissipative ergodic microstructures. Here, we concentrate rather on
spectroscopic problems.

It is clear from the above consideration that only escape to antennas and
inhomogeneous losses contribute to the fluctuating part
$\widetilde\Gamma_n=\Gamma_n-\Ghom$ of the widths. The latter are given now
by the imaginary parts of the complex eigenvalues of $\widetilde\Heff$.
Since $\widetilde\Heff$ is non-Hermitian, the eigenvalue problem
$\widetilde\Heff|n\rangle=\widetilde\mathcal{E}_n |n\rangle$ and
$\langle\tilde{n}|\widetilde\Heff=\langle\tilde{n}|\widetilde\mathcal{E}_n$
defines two sets of right and left eigenfunctions, which satisfy the
conditions of biorthogonality, $\langle\tilde{n}|m\rangle=\delta_{nm}$, and
completeness, $\sum_{n=1}^N|n\rangle\langle\tilde{n}|=1_N$. The matrix
$U_{nm}\equiv\langle n|m \rangle\neq\delta_{nm}$ differs from the unit one
and is known in nuclear physics as Bell-Steinberger nonorthogonality matrix
\cite{Bell1966} (see a compact description in \cite{Sokolov1989}). This
matrix features in two-point correlations in open systems seen, e.g., in
decay laws  \cite{Savin1997}. $U_{nn}$ appears also in optics via the
so-called Petermann factor of a lasing mode
\cite{Petermann1979,Siegman1989i,Schomerus2000a}. Some statistical aspects
of chaotic nonorthogonal eigenfunctions were recently studied in
\cite{Schomerus2000a,Chalker1998,Fyodorov2002ii}.

For a general non-Hermitian matrix left and right eigenvectors are
independent of each others. However, in our case $\widetilde\Heff$ is
complex symmetric (due to time-reversal invariance), implying that left
eigenfunctions are given by the transpose of the right ones,
$\langle\tilde{n}|=|n\rangle^{\mathrm{T}}$. As a result,
$\widetilde\Heff=\Psi\widetilde\mathcal{E}\Psi^{\mathrm{T}}$ can be
diagonalized by a \textit{complex} orthogonal transformation
\cite{Sokolov1989}, with
$\widetilde\mathcal{E}=\mathrm{diag}(\widetilde\mathcal{E}_1,\cdots,\widetilde\mathcal{E}_N)$
and $\Psi=(|1\rangle,\cdots,|N\rangle)$, that leads to the well-known pole
representation of the $S$-matrix, $S = 1 - i
A^{\mathrm{T}}\Psi(E_{\gamma}-\widetilde\mathcal{E})^{-1}\Psi^{\mathrm{T}}
A$, or in its components:
\begin{equation}\label{Spole}
S_{aa'} = \delta_{aa'} - i \sum_{n=1}^N \frac{\psi_n^a
\psi_n^{a'}}{E_{\gamma}-\widetilde\mathcal{E}_n}\,, \quad\qquad
\psi_n^a\equiv A^a|n\rangle=\sum_{k=1}^NA^a_{k}|n\rangle_{k}\,.
\end{equation}
The wave function component $\psi^a_n$ of the $n$-th mode excited through
the $a$-th channel is generally complex. This complexness is solely due to
biorthogonal nature of the eigenfunctions and is directly related to the
structure of the anti-Hermitian part of $\widetilde\Heff$. In particular,
all $\psi^a_n$ would be real were
$(\mathrm{Im}\widetilde\Heff)_{nm}\propto\delta_{nm}$ or, more generally, if
the anti-Hermitian part of the effective Hamiltonian commuted with its
Hermitian part\footnote{In this case both the Hermitian $H$ and
anti-Hermitian parts of $\Heff$ can be diagonalized simultaneously, thus the
eigenbasis being a conventional orthogonal one, as follows from
$H^{\mathrm{T}}=H$.}.

We proceed now with considering the case of tunneling coupling to antennas,
which was realized in experiment \cite{Barthelemy2005b}. This allows us to
neglect safely antenna contributions to $\Gamma_n$, approximating
$\widetilde\Heff\approx H-\frac{i}{2}CC^{\mathrm{T}}$. In the case of the
large but finite number $M_c$ of (contour) channels, the levels acquire on
average the width given by the so-called Weisskopf's estimate
\begin{equation}\label{Ginh}
\Ginh\equiv M_cT_c\frac{\Delta}{2\pi}\,,
\end{equation}
well-known in nuclear physics, see, e.g., \cite{Sokolov1989,Fyodorov1997}.
It is worth noting that this value can be formally linked to Sabine's law of
room acoustics, which determines the average width $\Gamma_{\mathrm{refl}}$
related to absorption at the cavity contour. One has
$\Gamma_{\mathrm{refl}}=cLT_c/(\pi S)$
\cite{Barthelemy2005b,Mortessagne1993}, where $S$ is the cavity area, $c$ is
the speed of light and $L$ is now the cavity perimeter. Making use of Weyl's
law for the mean level spacing $\Delta=c\lambda/S$ and putting
$M_c=L/(\lambda/2)$, we find that  $\Gamma_{\mathrm{refl}}$ is exactly
converted to $\Ginh$. This provides us with a further link between the
present model description and the microscopic treatment of
\cite{Barthelemy2005b,Barthelemy2005a} based on Maxwell's equations.

Fluctuations of the widths around $\Ginh$ are mostly due to those of the
matrix $CC^{\mathrm{T}}$. As follows from the central limit theorem,
fluctuations of off-diagonal matrix elements are suppressed as compared to
diagonal ones at $M_c\gg1$, $(CC^{\mathrm{T}})_{n\neq m}
\sim(CC^{\mathrm{T}})_{nn} / \sqrt{M_c}\sim\Ginh/\sqrt{M_c}$, so that they
contribute to $\widetilde\Gamma_n$ in the next-to-leading order in $1/M_c$.
However, off-diagonal matrix elements give a dominating contribution to the
mode nonorthogonality. Indeed, all essential features of the problem can be
most explicitly seen in the two-state approximation. Representing
$\widetilde\Heff$ as follows
\begin{equation}\label{2x2}
\widetilde\Heff\approx \left(\begin{array}{cc} \epsilon_1 & 0 \\ 0 &
\epsilon_2 \end{array}\right) - \frac{i}{2} \left(\begin{array}{cc}
\|C_1\|^2 & (C_1^{\mathrm{T}}C_2) \\ (C_2^{\mathrm{T}}C_1)& \|C_2\|^2
\end{array}\right)\,,
\end{equation}
where $C_{1,2}$ are the $M_c$-dimensional vectors of the corresponding
coupling amplitudes, one can easily solve the secular equation for its two
eigenvalues $\widetilde\mathcal{E}_{1,2}$, finding
$\widetilde\mathcal{E}_{1,2} = (\tilde\epsilon_1+\tilde\epsilon_2 \mp d)/2$,
where
$d=\sqrt{(\tilde\epsilon_1-\tilde\epsilon_2)^2-(C_1^{\mathrm{T}}C_2)^2}$ and
$\tilde\epsilon_{1,2}=\epsilon_{1,2}-\frac{i}{2}\|C_{1,2}\|^2$. The
corresponding eigenfunctions (we assume $\epsilon_1<\epsilon_2$) are given
by
\begin{equation}\label{basis}
|1\rangle = \mathcal{N} \left(\begin{array}{c} 1 \\ if\end{array}\right)
\quad \mathrm{and} \quad |2\rangle = \mathcal{N} \left(\begin{array}{c}  -if
\\ 1\end{array}\right)\,, \qquad
f=\frac{(C_1^{\mathrm{T}}C_2)}{\tilde\epsilon_2-\tilde\epsilon_1+d}\,,
\end{equation}
$\mathcal{N}^2=(1-f^2)^{-1}$ being the normalization constant. Then the
nonorthogonality matrix reads $U=|\mathcal{N}
|^2[(1+|f|^2)1_2+2\mathrm{Re}(f)\sigma_y]$, with the Pauli matrix $\sigma_y$.
The parameter $f$ controls the mode complexness, as follows from
\begin{equation}\label{psi}
\psi_{1,2}^a=\mathcal{N}(A_{1,2}^a\pm if A_{2,1}^a)
\end{equation}
for the mode wave functions.

In experiment \cite{Barthelemy2005b}, the complexness parameter
\begin{equation}
q^2=\langle(\mathrm{Im}\psi_n^a)^2\rangle/\langle(\mathrm{Re}\psi_n^a)^2\rangle
\end{equation}
was accessible only in the regime of the weak mode overlap due to
inhomogeneous losses (we stress, however, that the total width $\Ghom+\Ginh$
can be larger than $\Delta$). In this regime, $\Ginh\ll\Delta$, the
computation of $q$ can be easily carried out by making the use of the above
two-state approximation. One finds straightforwardly that
$q^2\approx\langle(\mathrm{Re}f)^2\rangle\approx\langle(C_1^{\mathrm{T}}C_2)^2\rangle/(2\Delta)^2$,
which determines now the average nonorhogonality matrix as $\langle
U_{nm}\rangle\approx(1+2q^2)\delta_{nm}$. The remaining Gaussian averaging
over random amplitudes (\ref{cov}) yields $\sum_{c,c'}\langle
C_1^cC_2^cC_1^{c'}C_2^{c'}\rangle = M_c(2\kappa_c\Delta/\pi)^2\approx
M_c(T_c\Delta/2\pi)^2$, as $T_c\approx4\kappa_c\ll1$ in our case of large
$M_c$ and fixed $\Ginh$. Collecting everything, we arrive at
\begin{equation}\label{q}
q\approx\frac{1}{\sqrt{M_c}}\frac{\Ginh}{2\Delta}
\end{equation}
that substantiates the proportionality between $q$ and $\Ginh$ established
experimentally \cite{Barthelemy2005b}. The mode complexness
(nonorthogonality) decreases as the number of absorptive channels increases,
in agreement with the general discussion presented above.

We discuss now the role of fluctuations in energy levels and contributions
from the other levels neglected so far. Restricting ourselves to the same
weak overlap regime, we can use the perturbation theory for wave functions
to find $q^2\approx\frac{1}{4}\sum_{m\neq
n}\langle(C_n^{\mathrm{T}}C_m)^2\rangle\langle(\epsilon_n-\epsilon_m)^{-2}\rangle$.
The known correlation function $R_2(\omega)$ of Gaussian orthogonal
ensembles, see \cite{Stoeckmann}, can be now used to get $\sum_{m\neq
n}\langle(\epsilon_n-\epsilon_m)^{-2}\rangle={\int}d\omega
\omega^{-2}R_2(\omega)$, which can be appreciated as (average square of) the
so-called ``level curvature'' studied in \cite{Fyodorov1995c,vonOppen1995}.
Since $R_2(\omega)\sim\omega$ as $\omega\to0$, this integral has a
logarithmic divergency  regularized by setting the lower limit $\sim\Ginh$
that also ensures us to stay within the perturbation theory. As a result,
(\ref{q}) is renormalized to yield a contribution $\sim\ln(\Delta/\Ginh)$.
One should expect that such a factor would be absent in a non-perturbative
treatment (see also the relevant discussion in \cite{Schomerus2000a}). The
computation of $q$ at arbitrary inhomogeneous absorption is still an
interesting open problem to consider.

In summary, we have presented the model description for inhomogeneous
(localized-in-space) losses in open chaotic systems and discussed thoroughly
the resulting complexness (or biorthogonality) of the mode wave functions.
In particular, the complexness parameter $q$ determining the relative weight
of the imaginary parts of the modes is found analytically to be proportional
to the inhomogeneous part $\Ginh$ of the widths in full agreement with
experimental results of \cite{Barthelemy2005b}.  Though the present
calculation of $q$ is perturbative in the small parameter $\Ginh/\Delta$, it
may nevertheless be valid in the intermediate or large modal overlap regime
where $\Ghom$ dominates, as, for instance, in the case of room acoustics or
elastodynamics \cite{Mortessagne1993, Kuhl2005r }. Our analysis may be also
relevant for problems of mode nonorthogonality outside of the scattering
systems as those considered recently in \cite{Chalker1998,Fyodorov2003}.

\acknowledgments %
We thank \textsc{Y.~V. Fyodorov} and \textsc{V.~V. Sokolov} for useful
comments. One of us (DVS) acknowledges gratefully the generous hospitality
of LPMC in Nice and the financial support of University of Nice during his
stay there. The financial support by the SFB/TR 12 of the DFG (\textsc{DVS})
is acknowledged with thanks.

%\bibliographystyle{epl}
%\bibliography{../Bib/refs,../Bib/books}

\begin{thebibliography}{25}

\bibitem{Kuhl2005r}
\Name{Kuhl~U., St{\"{o}}ckmann~H.-J., \and Weaver~R.}
\REVIEW {J. Phys. A: Math. Gen.}{38}{2005}{10433}.

\bibitem{Fyodorov2005r}
\Name{Fyodorov~Y.~V., Savin~D.~V., \and Sommers~H.-J.} %
\REVIEW {J. Phys. A: Math. Gen.}{38}{2005}{10731}.

\bibitem{Stoeckmann}
\Name{St{\"{o}}ckmann~H.-J.}
\Book{Quantum Chaos: An Introduction}
\Publ{Cambridge University Press, Cambridge, UK}
\Year{1999}.

\bibitem{Lobkis2000i}
\Name{Lobkis~O.~I. \and Weaver~R.~L.}
\REVIEW {J. Acoust. Soc. Am.}{108}{2000}{1480}.

\bibitem{Rozhkov2003}
\Name{Rozhkov~I., Fyodorov~Y.~V., \and Weaver~R.~L.}
\REVIEW {Phys. Rev. E}{68}{2003}{016204}.

\bibitem{Barthelemy2005b}
\Name{Barth{\'{e}}lemy~J., Legrand~O., \and Mortessagne~F.}
\REVIEW {Europhys. Lett.}{70}{2005}{162}.

\bibitem{Barthelemy2005a}
\Name{Barth{\'{e}}lemy~J., Legrand~O., \and Mortessagne~F.}
\REVIEW {Phys. Rev. E}{71}{2005}{016205}.

\bibitem{Brouwer2003}
\Name{Brouwer~P.~W.}
\REVIEW {Phys. Rev. E}{68}{2003}{046205}.

\bibitem{Kim2005}
\Name{Kim~Y.-H., Kuhl~U., St{\"o}ckmann~H.-J., \and Brouwer~P.~W.}
\REVIEW {Phys. Rev. Lett.}{94}{2005}{036804}.

\bibitem{Pnini1996}
\Name{Pnini~R. \and Shapiro~B.}
\REVIEW {Phys. Rev. E}{54}{1996}{R1032}.

\bibitem{Lewenkopf1992i}
\Name{Lewenkopf~C.~H., M{\"u}ller~A., \and Doron~E.}
\REVIEW {Phys. Rev. A}{45}{1992}{2635}.

\bibitem{Brouwer1997ii}
\Name{Brouwer~P.~W. \and Beenakker~C. W.~J.}
\REVIEW {Phys. Rev. B}{55}{1997}{4695}.

\bibitem{Verbaarschot1985}
\Name{Verbaarschot~J. J.~M., Weidenm{\"{u}}ller~H.~A., \and Zirnbauer~M.~R.}
\REVIEW {Phys. Rep.}{129}{1985}{367}.

\bibitem{Sokolov1989}
\Name{Sokolov~V.~V. \and Zelevinsky~V.~G.}
\REVIEW {Nucl. Phys. A}{504}{1989}{562}.

\bibitem{Fyodorov1997}
\Name{Fyodorov~Y.~V. \and Sommers~H.-J.}
\REVIEW {J. Math. Phys.}{38}{1997}{1918}.

\bibitem{Savin2003i}
\Name{Savin~D.~V. \and Sommers~H.-J.}
\REVIEW {Phys. Rev. E}{68}{2003}{036211}.

\bibitem{Rozhkov2004}
\Name{Rozhkov~I., Fyodorov~Y.~V., \and Weaver~R.~L.}
\REVIEW {Phys. Rev. E}{69}{2004}{036206}.

\bibitem{Bell1966}
\Name{Bell~J.~S. \and Steinberger~J.}
\Book{Proceedings of the Oxford International Conference on Elementary Particles,
  September, 1965}
\Publ{Rutherford High Energy Laboratory, Chilton, Berkshire, UK}
\Year{1966}
\Page{195}.

\bibitem{Savin1997}
\Name{Savin~D.~V. \and Sokolov~V.~V.}
\REVIEW {Phys. Rev. E}{56}{1997}{R4911}.

\bibitem{Petermann1979}
\Name{Petermann~K.}
\REVIEW {IEEE J. Quant. Electron.}{15}{1979}{566}.

\bibitem{Siegman1989i}
\Name{Siegman~A.~E.}
\REVIEW {Phys. Rev. A}{39}{1989}{1253}.

\bibitem{Schomerus2000a}
\Name{Schomerus~H., Frahm~K.~M., Patra~M., \and Beenakker~C. W.~J.}
\REVIEW {Physica A}{278}{2000}{469}.

\bibitem{Chalker1998}
\Name{Chalker~J.~T. \and Mehlig~B.}
\REVIEW {Phys. Rev. Lett.}{81}{1998}{3367}.

\bibitem{Fyodorov2002ii}
\Name{Fyodorov~Y.~V. \and Mehlig~B.}
\REVIEW {Phys. Rev. E}{66}{2002}{045202(R)}.

\bibitem{Mortessagne1993}
\Name{Mortessagne~F., Legrand~O., \and Sornette~D.}
\REVIEW {Chaos}{3}{1993}{529}.

\bibitem{Fyodorov1995c}
\Name{Fyodorov~Y.~V., \and Sommers~H.-J.} %
\REVIEW {Z. Phys. B}{99}{1995}{123}.

\bibitem{vonOppen1995}
\Name{von Oppen~F.} %
\REVIEW {Phys. Rev. E}{51}{1995}{2647}.

\bibitem{Fyodorov2003}
\Name{Fyodorov~Y.~V. \and Sommers~H.-J.}
\REVIEW {J. Phys. A: Math. Gen.}{36}{2003}{3303}.

\end{thebibliography}
%\end{document}

\end{document}